\documentclass[conference]{IEEEtran}
\IEEEoverridecommandlockouts
\usepackage{cite}
\usepackage{amsmath,amssymb,amsfonts}
\usepackage{algorithmic}
\usepackage{graphicx}
\usepackage{textcomp}
\usepackage{xcolor}
\usepackage{tikz}
\usepackage{multirow}
\newcommand*\circled[1]{\tikz[baseline=(char.base)]{\node[shape=circle,fill,inner sep=0.2pt] (char){\textcolor{white}{#1}};}}

\def\BibTeX{{\rm B\kern-.05em{\sc i\kern-.025em b}\kern-.08em
    T\kern-.1667em\lower.7ex\hbox{E}\kern-.125emX}}
\begin{document}

\title{Analysis and Optimization of I/O Cache Coherency Strategies for SoC-FPGA Device}
\author{\IEEEauthorblockN{Seung Won Min}
\IEEEauthorblockA{\textit{Electrical and Computer Engineering} \\
\textit{University of Illinois}\\
Urbana, USA \\
min16@illinois.edu} \\
\IEEEauthorblockN{Jinjun Xiong}
\IEEEauthorblockA{
\textit{IBM T.J. Watson Research Center}\\
Yorktown Heights, USA \\
jinjun@us.ibm.com}
\and
\IEEEauthorblockN{Sitao Huang}
\IEEEauthorblockA{\textit{Electrical and Computer Engineering} \\
\textit{University of Illinois}\\
Urbana, USA \\
shuang91@illinois.edu}\\
\IEEEauthorblockN{Deming Chen}
\IEEEauthorblockA{\textit{Electrical and Computer Engineering} \\
\textit{University of Illinois}\\
Urbana, USA \\
dchen@illinois.edu}
\and
\IEEEauthorblockN{Mohamed El-Hadedy}
\IEEEauthorblockA{\textit{Electrical and Computer Engineering} \\
\textit{California State Polytechnic University}\\
Ponoma, USA \\
mealy@cpp.edu} \\
\IEEEauthorblockN{Wen-mei Hwu}
\IEEEauthorblockA{\textit{Electrical and Computer Engineering} \\
\textit{University of Illinois}\\
Urbana, USA \\
w-hwu@illinois.edu}
}

\maketitle

\begin{abstract}
%
%
%
%
%
%

Unlike traditional PCIe-based FPGA accelerators, heterogeneous SoC-FPGA devices provide  tighter integrations between software running on CPUs and hardware accelerators. 
Modern heterogeneous SoC-FPGA platforms support multiple I/O cache coherence options between CPUs and FPGAs, but these options can have inadvertent effects on the achieved bandwidths depending on applications and data access patterns. 
To provide the most efficient communications between CPUs and accelerators, understanding the data transaction behaviors and selecting the right I/O cache coherence method is essential.
In this paper, we use Xilinx Zynq UltraScale+ as the SoC platform to show how certain I/O cache coherence method can perform better or worse in different situations, ultimately affecting the overall accelerator performances as well.
Based on our analysis, we further explore possible software and hardware modifications to improve the I/O performances with different I/O cache coherence options.
With our proposed modifications, the overall performance of SoC design can be averagely improved by 20\%.

\end{abstract}

\begin{IEEEkeywords}
FPGA, heterogeneous computing, cache, cache coherence
\end{IEEEkeywords}

\section{Introduction}
\label{section:introduction}
Heterogeneous SoC-FPGA platforms such as Xilinx Zynq UltraScale+ MPSoC provide flexible development environment with tightly-coupled interfaces between different processing units inside.
Depending on the needs of users, these processing units can be combined and programmed to provide the most suitable configuration.
For the different components to operate seamlessly together, it is important to understand how data coherency between them are managed.
For the traditional server or desktop class machines, there is little meaning of configuring the host system's I/O cache coherence for general FPGA designers because often: 1) manufacturers do not provide any documentations of that level of detail or 2) I/O cache coherence is enabled by default in such scales of systems.
On the other hand, in SoC-FPGA design, all available I/O cache coherence options are fully disclosed to the FPGA designers and the designers are responsible of choosing the most suitable methods for target applications.

However, choosing the right I/O cache coherence method for different applications is a challenging task because of it's versatility.
By choosing different methods, they can introduce different types of overheads.
Depending on data access patterns, those overheads can be amplified or diminished.
In our experiments, we find using different I/O cache coherence methods can vary overall application execution times at most 3.39$\times$.
This versatility not only makes designers hard to decide which methods to use, but also can mislead them to wrong decisions if performance evaluations are incomprehensive.
SoC IP providers such as Xilinx and ARM provide high-level guides~\cite{arma53,zynqus+} of using different I/O cache coherence methods and interfaces, but these are often vague and do not include any quantitative analysis.

In this work, we analyze the effects of using different I/O cache coherence methods in SoC-FPGA as detail as possible and provide general guide of using each method.
Our I/O cache coherence performance analysis consists of two parts: software costs and hardware costs.
The software cost denotes how much the software portion of applications can be affected to maintain certain types of I/O cache coherence methods.
The hardware cost denotes how much the hardware complexities added to maintain I/O cache coherence can affect I/O bandwidths.
Later in this paper, both of the costs are combined to evaluate the total cost of I/O cache coherence.
%
%
%
%
%
%
%
Throughout the experiments, we use Xilinx's Zynq UltraScale+ platform which supports variety of interface options including hardware coherent I/O bus and direct accesses to L2 cache.
The contributions of this paper can be summarized as follows:
\begin{itemize}
    \item Evaluate software and hardware costs of using different I/O cache coherence methods.
    \item Introduce several optimization techniques which can eliminate some I/O cache coherence costs.    
    \item Provide a complete guide of achieving efficient I/O cache coherence based on real hardware evaluation results.
\end{itemize}

The rest of the paper is organized as follows.
In Section~\ref{section:background}, we explain backgrounds of different I/O cache coherence strategies in detail.
In Section~\ref{section:test_environment}, we elaborate our experiment environment.
In Section~\ref{section:cache_and_fpga}, we show our software and hardware I/O cache coherence cost evaluation results.
In Section~\ref{section:optimize}, we provide a general guide of I/O cache coherence optimizations.
Section~\ref{section:related} discusses related works.
Finally in Section~\ref{section:conclusion}, we summarize our work and conclude this paper.

\section{I/O Cache Coherence}
\label{section:background}

In a modern system design, it is common to use memory as a shared buffer to transfer data between CPUs and I/O devices~\cite{berg2009maintaining}.
However, with CPU caches, it is possible the data inside the shared buffer is physically scattered over the caches and DRAM.
In such case, depending on the perspective, the buffer may contain different values.
To avoid the situation, I/O cache coherence is required to maintain data coherency and consistency between CPUs and I/O devices.
I/O cache coherence can be achieved in several ways.
First, certain regions of memory can be disabled from caching.
Second, CPUs can manually flush or invalidate cache lines before any I/O data transactions.
Third, hardware implementations can be added so I/O devices can snoop CPU caches.
In this section, we describe the methods and briefly discuss their benefits and costs.

\subsection{Allocating Non-cacheable Memory}

The simplest way of achieving I/O cache coherency is making memory accesses non-cacheable.
This does not need to be enforced globally, and it can be narrowed down to specific memory regions which are shared between CPUs and I/O devices by setting appropriate ISA-dependent virtual page attributes.
However, in this case, CPU memory accesses to the regions may lose benefits of data locality.

\subsection{Software I/O Coherency}
Software I/O coherency requires CPUs to manually flush or invalidate cache lines by executing cache maintenance instructions before any data transactions between CPUs and I/O devices are made.
In this method, CPUs can still cache data from the memory regions shared with I/O devices, but the manual cache instructions are in critical paths of I/O data transactions and it can decrease effective bandwidths~\cite{loghi2006cache}.
Furthermore, global memory fences should be inserted between the cache instructions and data accesses to guarantee no data accesses precedes the cache instructions.

\subsection{Hardware I/O Coherency}
Hardware coherency relies on hardware implementations included in host systems which let I/O devices to snoop CPU caches.
This I/O coherence method requires the least amount of software designers' attentions as the shared memory regions can be treated as cacheable, and it does not require any cache maintenance instructions.
Achieving the cache snooping can be largely done in two ways.
First, I/O buses between CPUs and I/O devices can be modified so every memory access requests from I/O devices cause cache snoop requests as well.
Depending on snooping results, I/O devices can directly grab data from caches for readings or automatically invalidate stale cache lines from CPU caches when writing to memory.
However, resolving cache snoop requests may require several extra bus cycles between different memory requests which can reduce I/O bandwidths~\cite{girao2007cache}.
The second way is directly connecting I/O devices to caches.
In this case, I/O devices generate cache snooping requests like other CPU cores.
The difference compared to the first method is in this case, I/O data requests are treated as regular CPU data requests and each request generates a cache line allocation.
This would be beneficial if the cache line allocated is reused frequently, but with inappropriate data access patterns it can end up evicting useful cache lines for CPUs.

\section{Experiment Environment}
\label{section:test_environment}

\begin{figure}[t]
\centerline{\includegraphics[width=0.8\linewidth]{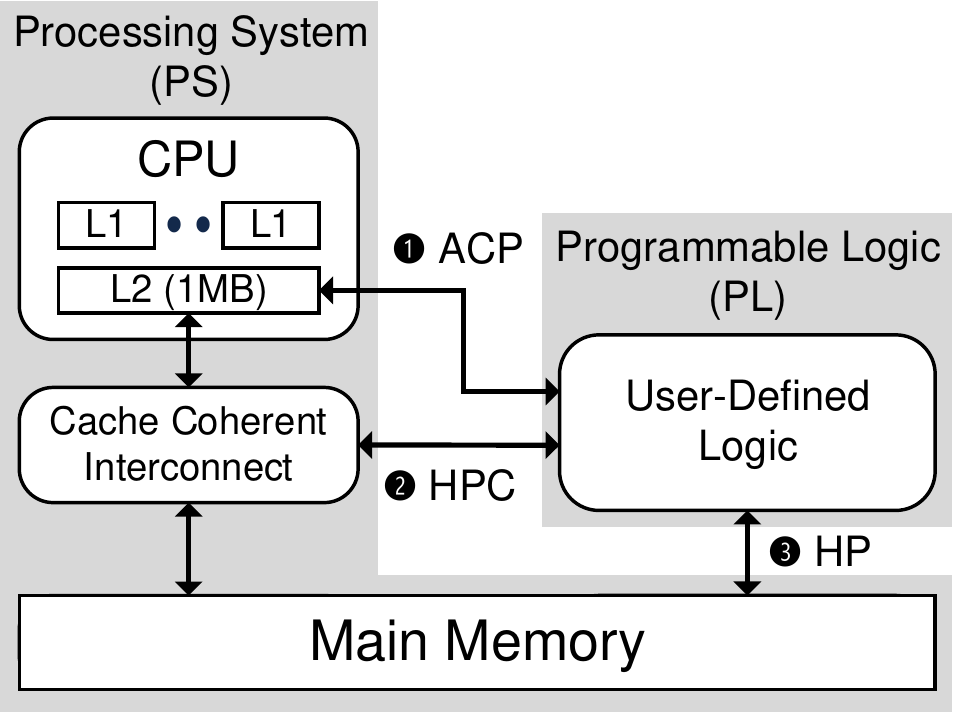}}
\caption{Simplified block diagram of possible I/O configurations in Xilinx Zynq UltraScale+. \protect\circled{1} Accelerator Coherency Port (ACP) can access L2 cache directly. \protect\circled{2} High Performance Coherent (HPC) interface goes through coherent I/O bus where it can issue cache snooping requests to CPU cache. \protect\circled{3} High Performance (HP) interface goes to memory directly and I/O cache coherence should be dealt by CPU.}
\label{fig:zynq}
\vspace{-12pt}
\end{figure}

\setlength\tabcolsep{3pt}
\begin{table}[b]
\caption{Available PL Interfaces and Data Coherency Methods in Zynq UltraScale+}
\begin{center}
\begin{tabular}{|c|c|c|c|c|}
\hline
\multirow{2}{*}{Alias} & \multirow{2}{*}{Inteface} & Memory & Data channel & Coherency\\
& & Allocation & is connected to & Method\\
\hline
HP (NC) & HP & Non-cacheable & Memory & Not Required\\
\hline
HP (C) & HP & Cacheable & Memory & Cache Inst.\\
\hline
\multirow{2}{*}{HPC} & \multirow{2}{*}{HPC} & \multirow{2}{*}{Cacheable} & Memory \& & \multirow{2}{*}{H/W Coherent}\\
& & & Cache (Read-only) & \\
\hline
ACP & ACP & Cacheable & Cache & H/W Coherent\\
\hline
\end{tabular}
\label{table:io_match}
\end{center}
\end{table}

All experiments in this paper are done based on Xilinx Zynq UltraScale+ MPSoC.
Zynq Ultrascale+ has Processing System (PS) block and Programmable Logic (PL) block as described in Fig.~\ref{fig:zynq}. 
PS consists of hard IPs such as CPU, coherent I/O, and memory.
PL consists of programmable logic and can be programmed by users like regular FPGAs.
Between the two blocks, there are several types of I/O available.
\circled{1} Accelerator Coherency Port (ACP) interface can access shared L2 cache (1MB) directly.
However, this port does not fully comply with Advanced eXtensible Interface 4 (AXI4) protocol which is commonly used in Xilinx IPs. 
Since there is no publicly available ACP adapter IP, we developed ACP$\leftrightarrow$AXI4 converter for our experiments.
\circled{2} High Performance Coherent (HPC) interface goes through coherent I/O bus where it can issue cache snooping requests to the CPU cache.

ARM Cache Coherent Interconnect 400 (CCI-400)~\cite{armcci} is used for this coherent I/O bus and it uses AXI Coherency Extensions (ACE) and ACE-Lite protocols to support cache coherency.
ACE protocol supports bi-directional (Cache$\leftrightarrow$Cache) cache coherency and ACE-Lite supports one-directional (Device$\rightarrow$Cache) cache coherency.
CCI-400 can support up to two ACE ports where the one is already occupied by ARM Cortex-A53 CPU.
We do not use the other ACE port in this experiment since our accelerators do not implement any private caches.
In context of Zynq UltraScale+, HPC interfaces are only using ACE-Lite protocols.
\circled{3} High Performance (HP) interface goes to memory directly and I/O cache coherence should be dealt by the CPU.
All interfaces are 128-bit wide and we fix interface frequencies to 300 MHz throughout our experiments, providing the maximum theoretical bandwidths of 4.8 GB/s.
Table~\ref{table:io_match} summarizes overall Zynq UltraScale+ interfaces and possible I/O cache coherence methods.
In the rest of the paper, we refer the HP interface with non-cacheable and cacheable memory allocations as HP (NC) and HP (C), respectively.

Software I/O coherency implementation is embedded in Xilinx drivers and the drivers are capable of identifying the buffer allocation types.
If the buffers are non-cacheable, the drivers do not manually flush or invalidate caches.
If the buffers are cacheable, the drivers automatically perform cache flushes and invalidations.

\begin{figure}[t]
\centerline{\includegraphics[width=\linewidth]{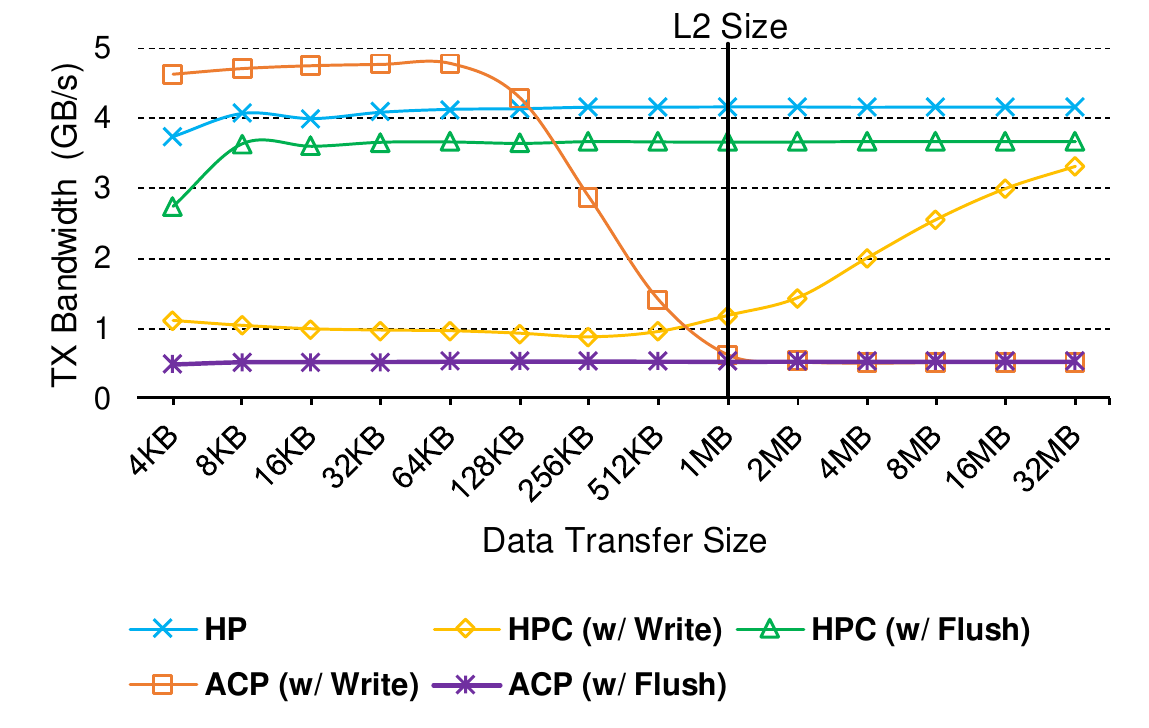}}
\vspace{-6pt}
\caption{I/O bus TX (CPU$\rightarrow$PL) bandwidth comparison. No software overhead is included in this measurement.}
\label{fig:tx}
\vspace{-12pt}
\end{figure}

\section{I/O Cache Coherence and SoC-FPGA}
\label{section:cache_and_fpga}

In this section, we evaluate hardware and software costs of different I/O cache coherence methods.
For the hardware cost, we are interested in identifying how much the extra steps required to resolve cache snoop requests in hardware can negatively affect I/O bandwidths.
For the software cost evaluation, we measure CPU overheads added when hardware coherent I/O interfaces are not supported.

\begin{figure}[t]
\centerline{\includegraphics[width=\linewidth]{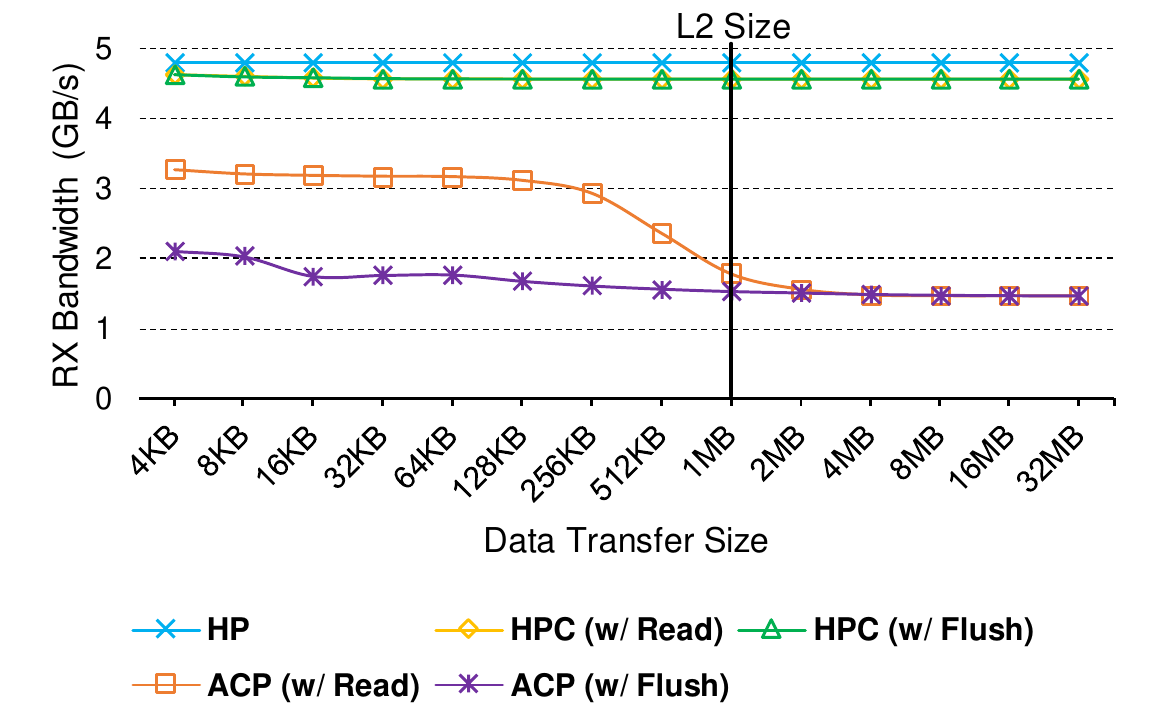}}
\vspace{-6pt}
\caption{I/O bus RX (PL$\rightarrow$CPU) bandwidth comparison. No software overhead is included in this measurement.}
\label{fig:rx}
\vspace{-12pt}
\end{figure}

\subsection{Hardware Cost Evaluation}
\label{section:io_bandwidth}

In this experiment, we measure raw bandwidths of non-hardware coherent I/O (HP) and hardware coherent I/O (HPC and ACP) interfaces.
The raw bandwidth here means the pure interface bandwidths without any software overheads included.
To measure CPU to PL (TX) and PL to CPU (RX) bandwidths, we program PL to initiate data transfers and count how many bus clock cycles spent. 
For the hardware coherent I/Os, we'd like to also know if there are any bandwidth differences when the shared buffer data for both TX and RX cases are cached or not.
To achieve this, we intentionally read/write or flush the entire range of the shared buffers before the data transfers begin.
The summary of the test setups can be found at Table~\ref{table:io_test_setup}.
We do not differentiate between HP (NC) and HP (C) in this experiment as their differences are only at software costs.

\setlength\tabcolsep{5pt}
\begin{table}[b]
\caption{Raw Bandwidth Test Setup}
\begin{center}
\begin{tabular}{|c|c|c|}
\hline
\multirow{2}{*}{Direction} & \multirow{2}{*}{Interface} & Before data transfer\\
& & the buffer has been\\
\hline
 & HP & \\
CPU & HPC (w/ Write) & Written\\
$\downarrow$ & HPC (w/ Flush) & Flushed\\
PL & ACP (w/ Write) & Written\\
 & ACP (w/ Flush) & Flushed\\
\hline
 & HP & \\
PL & HPC (w/ Read) & Read\\
$\downarrow$ & HPC (w/ Flush) & Flushed\\
CPU & ACP (w/ Read) & Read\\
& ACP (w/ Flush) & Flushed\\
\hline
\end{tabular}
\label{table:io_test_setup}
\end{center}
\end{table}

Fig.~\ref{fig:tx} shows the TX bandwidth measurement results.
Starting from the HP results, we observe almost no differences in TX bandwidths while sweeping from 4KB to 32MB data transfers.
There is a small bandwidth drop at 4KB due to the initial DRAM access latency, but the overhead of the latency becomes almost not visible as the data transfer size increases.

In case of HPC, we see huge differences when the data is cached or not.
For HPC (w/ Flush), there is only a small bandwidth drop compared to HP, but for HPC (w/ Write), the TX bandwidth decreases significantly.
Based on this analysis, we can assume the data flow path from CPU cache to the device is sub-optimal in Zynq UltraScale+.
Writing larger amount of data to the buffer attenuates this problem as the maximum amount of cached data is limited by the L2 size.
Still, to reach near the peak HPC bandwidth, more than 32 MB of data should be transferred.

ACP bandwidth is nearly reaching 4.8 GB/s with small sizes of data, but it starts to sharply drop as the data size approaches toward the L2 size.
A53 L2 cache does not have hardware prefetching unit and therefore all cache accesses without pre-populated cache lines need to pay cache miss penalties.
By observing the measurement results, we can assume writing more than 64KB of data in one time starts to evict its previously allocated cache lines.
Currently, A53 L2 cache is using random cache replacement policy, but future SoC-FPGA platforms using least recently used cache replacement policy may push back the self eviction point.
When the buffer is completely flushed before the data transfer, ACP constantly suffers from the low bandwidth as all cache accesses cause cache misses.

For the RX bandwidth measurement results, we do not see any significant bandwidth changes beside ACP.
In Fig.~\ref{fig:rx}, both HP and HPC are reaching near 4.8 GB/s of bandwidths in all cases.
In case of ACP, we observe a similar trend to the TX case where the ACP bandwidth is higher when most of the data are cached.

The bandwidth discrepancies between the RX and TX can be due to the cache coherency protocol.
For example, Molka et al.~\cite{molka2009memory} describes different cache read and write bandwidths in the Intel's Nehalem processors due to the cache coherency protocol used in them.

\begin{figure}[t]
\centerline{\includegraphics[width=\linewidth]{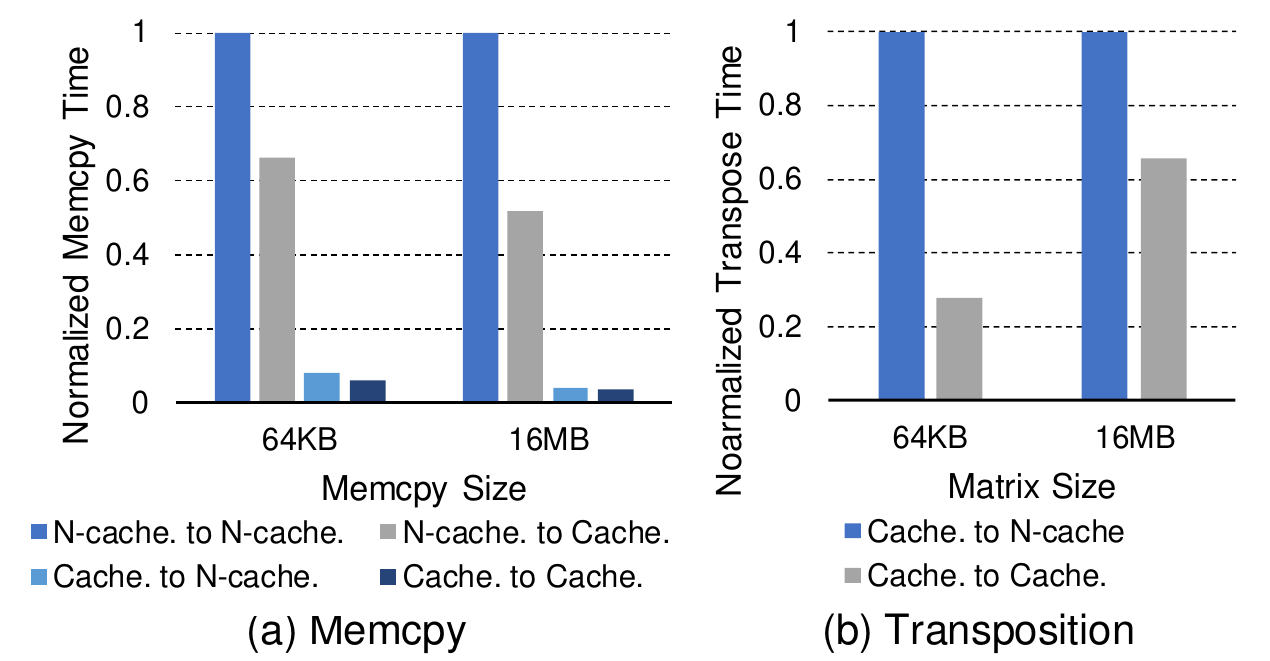}}
\caption{(a) Memcpy execution time comparison using different combinations of non-cacheable and cacheable source/destination buffers. (b) Matrix transpose execution time comparison with non-cacheable and cacheable destination buffers.}
\label{fig:noncacheable_memcpy}
\vspace{-12pt}
\end{figure}

\subsection{Software Cost Evaluation}
\label{section:cpu_overhead}

In this section, we evaluate non-cacheable memory access bandwidths and manual cache operation costs.
The advantage of using caches is well evaluated in the past~\cite{jouppi1990improving,lam1991cache}, but we include the evaluation in this paper for the completeness of I/O cache coherency evaluation.
For the non-cacheable memory access evaluation, we first measure four types of memory copy operations: non-cacheable to non-cacheable, non-cacheable to cacheable, cacheable to non-cacheable, and cacheable to cacheable.
All memory copies are done using \texttt{memcpy()} function from the C library.
In Fig.~\ref{fig:noncacheable_memcpy} (a), we find the bandwidth penalty is as large as 30$\times$ when reading from the non-cacheable region compared to reading from the cacheable region.
On the other hand, the memory writes to the non-cacheable regions remains almost the same because the Write-Combine (WC) function can combine multiple non-cacheable write requests to a single larger memory write.
This feature will be further discussed in Section~\ref{section:hardware_exploit}.
%
%
%

Still, the WC is only active in regular memory access patterns and CPUs can suffer from long memory latencies with irregular memory write patterns.
In Fig.~\ref{fig:noncacheable_memcpy} (b), we measure execution times of matrix transpositions to different types of memory.
In this experiment, the source matrix is stored in cacheable memory region and the destination for the transposed matrix is located in non-cacheable memory region.
When the entire matrix can fit in the cache, the cacheable memory is about 4$\times$ faster than the non-cacheable memory.
When the matrix size is much larger than the cache size, the cacheable memory is still about 1.33$\times$ faster than the non-cacheable memory.

When manual cache instructions are needed, the CPU overhead added heavily depends on other CPU workloads and the total number of buffers flushed or invalidated.
In Linux, after each buffer is flushed or invalidated, global memory barrier should be inserted to guarantee no memory accesses are reordered.
If this global memory barrier needs be executed multiple times while heavy memory accesses are being made, the overall CPU performance can be severely degraded.

\begin{figure}[t]
\centerline{\includegraphics[width=\linewidth]{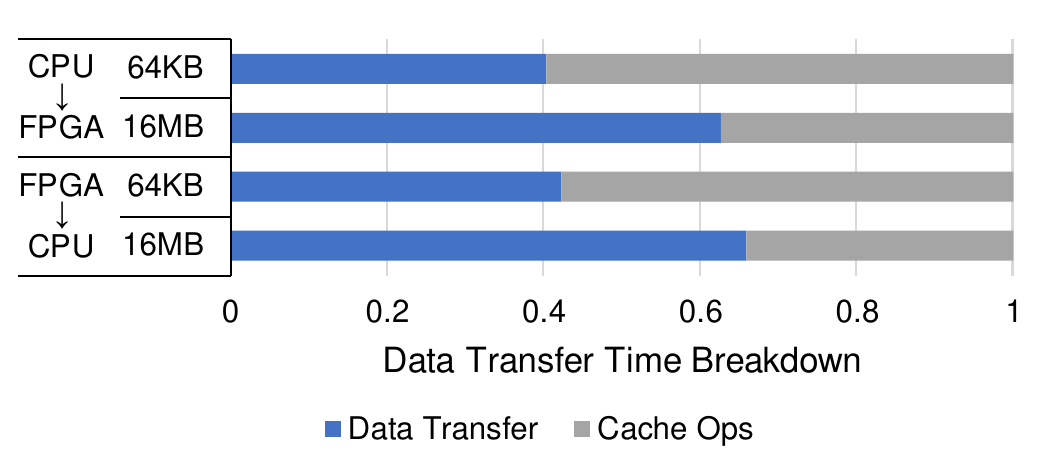}}
\caption{Data transfer time breakdown with manual cache maintenance instructions.}
\label{fig:data_move_cache}
\end{figure}

In Fig.~\ref{fig:data_move_cache}, we show data transfer time breakdown with manual cache instructions.
With a smaller data size, the manual cache instructions take majority of the total data transfer time.
With a larger data size, the overhead of the memory barrier takes less portion of the total data transfer time and the total overhead of the manual cache instructions become smaller.
We find the directions of the data transfers do not significantly affect the manual cache instruction overheads.

\begin{figure*}[t]
\centerline{\includegraphics[width=7in]{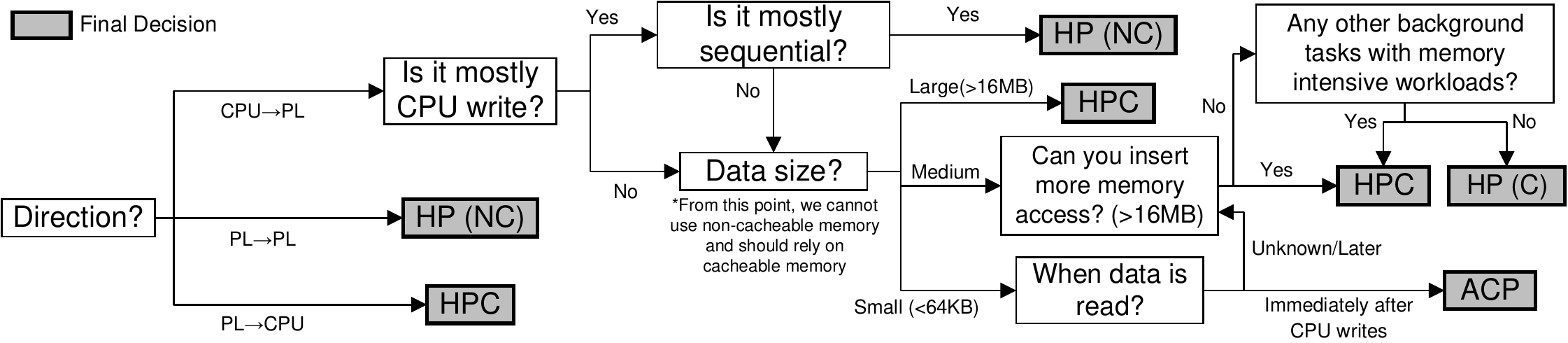}}
\caption{Decision tree for selecting the optimal I/O cache coherence method.}
\label{fig:decision_tree}
\vspace{-12pt}
\end{figure*}

\section{Optimizing Data Transactions}
\label{section:optimize}
In this section, we suggest several I/O cache coherence optimization techniques to achieve the most effective data transaction behaviors.
First, we introduce several hardware features which can be exploited to remove some I/O cache coherence overheads.
Second, we present a decision tree (Fig.~\ref{fig:decision_tree}) which can be utilized to optimize I/O cache coherence selections.
Finally, we apply our decision tree to several applications and compare the overall performances with baseline designs.

\subsection{Exploiting Hardware Features}
\label{section:hardware_exploit}

\subsubsection{Wribe Combine (WC)}
WC is a cache feature which can combine multiple write accesses to non-cacheable regions into a single larger memory write request~\cite{benkual2003system}.
Compared to requesting multiple small memory writes, requesting a single larger memory write can better utilize the memory bandwidth.
To activate this feature, consecutive write requests should be contiguous in memory address space in certain degree.
The minimum requirement for the contiguity may depend on CPU architecture, and A53 requires at least the write requests are 128-bit aligned.
For example, if there are four integer (4-byte) write requests to address of 0x00, 0x01, 0x02 and 0x03, then they can be combined into a single 128-bit write request.
When the write requests are pointing to different memory addresses resulting into different memory alignments, they need to be split into different memory write requests.

\subsubsection{Cache Bypass}
In Section~\ref{section:io_bandwidth}, we showed the CPU$\rightarrow$PL bandwidth of HPC interface can be significantly lower when the data is cached.
It is possible to resolve this by manually flushing cache lines, but this costs CPU cycles in exchange.
One way to implicitly flush the cache lines is using cache bypass function in hardware~\cite{johnson1999run}.
Cache bypass can be used in cacheable memory region where caches decide not to allocate certain cache lines for certain data access patterns.
In A53, similar function, called \textit{Read Allocate Mode}, is implemented to not allocate cache lines when there is a massive amount of writes with regular access patterns.
This kind of behavior can be often observed when using \texttt{memset()}.
With this feature, without explicitly executing cache flush instructions, data can be directly written into DRAM even if the memory regions are cacheable.
However, if the memory writes are done with irregular patterns, the read allocate mode is not activated.

\subsection{I/O Cache Coherence Decision Tree}

Gathering all explorations from previous sections, we build a decision tree (Fig.~\ref{fig:decision_tree}) to provide a general I/O cache coherence optimization flow.
The total cost of I/O cache coherence can be roughly estimated as follows:
\[ (total \: cost) = \frac{\alpha}{(raw \: bandwidth)} + (software \: cost) \]
Here, the $\alpha$ represents the bandwidth requirement of an application.
We first categorize all data transaction types into CPU to PL, PL to PL, and PL to CPU.
Just to clarify, in this decision tree, we are only accounting to the cases where a shared memory (mostly host DRAM) between two instances is used as a data communication medium.
Without any shared memory, there are no I/O cache coherency issues.
Our decision tree strategy focuses on minimizing unexpected risks rather than maximizing possible gains.
The parameter values set in this decision tree can be rather conservative.

For the communication between PL logics, there is no CPU involvement and therefore using HP (NC) is the best.
For the PL to CPU case, we conclude using HPC interface is the best in general as it can provide relatively high memory bandwidth while not introducing additional software costs.
The memory bandwidth loss with the HPC interface in this case compared to the HP is about 5\% (Fig.~\ref{fig:rx}).

CPU to PL case is more complex than the former two cases as the raw bandwidth differences are huge in this case.
In this case, we first check if the TX buffer is mostly used for CPU write.
If the CPU is mostly writing to the buffer, then we check if the writing is mostly done in sequential manner.
If the memory write patterns are sequential or can be modified to be sequential, then we can safely use the non-cacheable memory allocation.
If the writes cannot be made sequential or the CPU needs to make substantial amount of read requests from this buffer, the buffer cannot be made non-cacheable.
From this point, we need to rely on HP (C), HPC, or ACP.

Using HP (C) is discouraged in general since executing extra cache instructions and memory barriers can only have negative affects in terms of performances.
To use HPC or ACP, we must check how much of the data to be transferred is cached as the raw bandwidths of HPC and ACP vary a lot depending on the data locations.
However, because it is impossible to know the exact location of data before we access the cache, we rely on several intellectual guesses.
First, we check the size of the data.
If the data size is large enough ($>$16MB), based on our observation from Fig.~\ref{fig:rx}, we can obtain relatively high bandwidth with HPC.
Second, if the data size is small ($<$64KB) and the accelerator reads the data immediately after the CPU writing, we can use ACP to maximize the bandwidth.
Third, if none of the above cases were true, we can consider reordering some other workloads to just before the accelerator data reads.
For example, in video streaming, we can add some delay of several frames to make latter frames to evict former frames from the cache.
If the reordered workloads can make large enough amount of memory accesses ($>$16MB), most of the data will be evicted from the cache and we can use HPC.
If this is also impossible, then we finally need to consider using HP (C).
Before we choose HP (C), one thing we may need to consider is if there are any background tasks which are memory intensive.
If there are any such tasks, we should still consider using HPC as memory barriers inserted by HP (C) is likely to slow down the overall CPU performances.

\begin{figure}[t]
\centerline{\includegraphics[width=\linewidth]{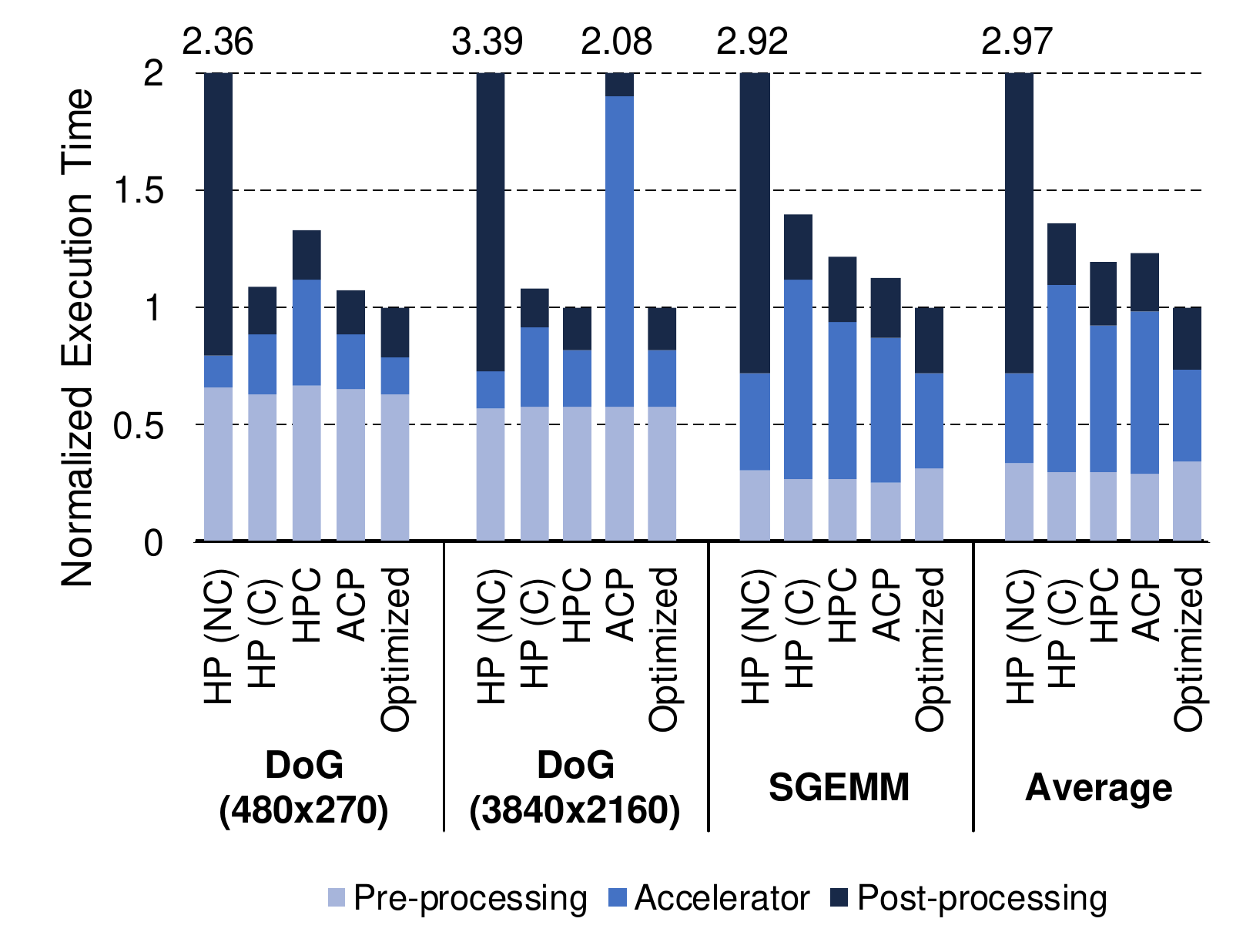}}
\caption{Benchmark results with using different I/O cache coherence methods. Difference of Gaussian (DoG) is tested with different image sizes.}
\label{fig:eval}
\vspace{-12pt}
\end{figure}

\subsection{Case-Study Evaluations}
To evaluate our decision tree, we use modified \textit{Difference of Gaussian (DoG)} filter from \textit{xfOpenCV}~\cite{xfopencv}, \textit{SGEMM}, and \textit{CHaiDNN}~\cite{chaidnn} with \textit{AlexNet} as case-study examples.
All applications are written in C++ and synthesized with Xilinx SDSoC.
DoG takes grayscale images as inputs and generates two outputs.
The first output is generated by directly applying a gaussian filter to the input and the second output is generated by passing the first output to another gaussian filter.
Later, two output images are subtracted to each other and the final output is generated.
This difference of the gaussian filtered images are often used for edge detections.
For this application, we use CPU to convert RGB images to grayscale images and subtract two gaussian filtered images.
Accelerator is used for accelerating the gaussian filters.
For SGEMM, we implement a 128$\times$128 matrix multiplication accelerator and perform block matrix multiplication for larger input matrices.
CPU is responsible of cropping input matrices into 128$\times$128 blocks and feeding into the SGEMM accelerator and accumulating the accelerator outputs into the output matrix.
CHaiDNN accelerates convolution and pooling layers of DNN and CPU is responsible of quantizing input images and de-quantizing accelerator outputs.

For the baselines, we implement designs with pure HP (NC), HP (C), HPC, or ACP options.
Due to the design complexity, we only compare between HP (NC), HP (C), and optimized version for CHaiDNN.
The baseline CHaiDNN design from Xilinx only uses HP (NC) and HP (C).
The optimized designs follow the decision tree we created.
The modifications are only done in memory allocation types and interface connections and accelerators are not modified while comparing with other I/O cache coherence methods.

Fig.~\ref{fig:eval} shows the benchmark results of DoG with different image sizes and SGEMM.
In average, our optimized version achieved at least 20\% of execution time reduction compared to any other baseline configurations.
In general, HP (NC) has the smallest accelerator execution times due to is high raw bandwidth, but the post-processing times have been greatly increased.
HP (C) in general has very long accelerator execution times because of manual cache instructions and memory barriers.
HPC performs well when the input sizes are large, but starts to suffer from low raw bandwidth when the inputs are small due to the reason explained in Section~\ref{section:io_bandwidth}.
In opposite, ACP performs well when the input sizes are small, but as the input sizes increase the cache hit rates become lower and the accelerator execution times start to skyrocket.

\begin{figure}[t]
\centerline{\includegraphics[width=\linewidth]{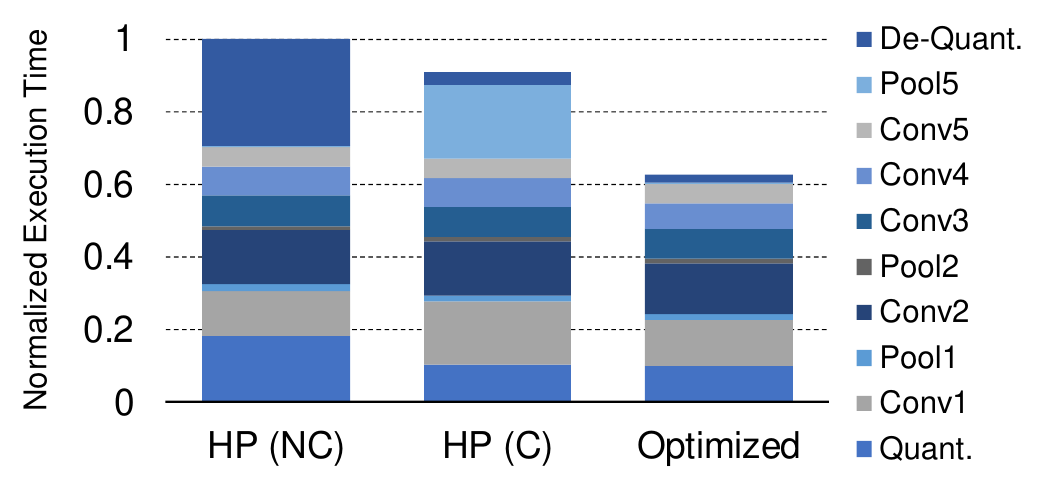}}
\caption{Benchmark results of CHaiDNN with different I/O cache coherence methods. Quantizations and de-quantizations are done at CPU. The execution order is from bottom to top (Quant.$\rightarrow$Conv1$\rightarrow$...$\rightarrow$Pool5$\rightarrow$De-Quant.).}
\label{fig:alexnet}
\vspace{-12pt}
\end{figure}

Fig.~\ref{fig:alexnet} shows the AlexNet execution time breakdown with CHaiDNN.
HP (NC) greatly suffers from non-cacheable memory accesses during both quantizations and de-quantizations.
HP (C) has slightly better performance than HP (NC), but still need to spend non-negligible amount of time executing manual cache instructions.
The optimized version removes the penalties of both HP (NC) and HP (C) and reduces the execution time by 37.2\% and 30.9\% compared to HP (NC) and HP (C), respectively.

\section{Related Works}
\label{section:related}

There are several I/O cache coherence bandwidth researches with older SoC-FPGA platforms such as Xilinx's Zynq-7000 and Altera's Cyclone V~\cite{powell2015statistical,silva2015comparison,sadri2013energy,vogel2015evaluation,sklyarov2015analysis,molanes2018performance,molanes2015characterization}.
For both platforms, the only available hardware coherent I/O port is ACP.
\cite{silva2015comparison,sadri2013energy,vogel2015evaluation,sklyarov2015analysis,molanes2015characterization,molanes2018performance} are limited to evaluating raw I/O bandwidths of using different ports and did not include software cost evaluations.
\cite{powell2015statistical} has evaluated software costs of I/O cache coherence, but only with a fixed data access pattern.
 
\section{Conclusion}
\label{section:conclusion}
The costs of different I/O cache coherence methods varies widely depending on applications.
Approaching the I/O cache coherence optimization problem should be done in bottom-up fashion including both software and hardware profilings.
In this paper, we presented multiple I/O cache coherence methods of SoC-FPGA and optimization techniques based on thorough analysis of Zynq UltraScale+ platform.
By properly combining different I/O cache coherence methods, we showed the overall execution time can be reduced by 20\%.
In this paper, we mainly discussed the I/O cache coherence in a context of CPU-to-accelerator connections, but this can be also applied to other device connections such as high-speed Ethernet, GPU, and NVMe.
Considering that modern SoC-FPGA platforms can support many different kinds of peripherals, a good understanding of I/O cache coherence optimization will be more important in the future.

\section{Acknowledgements}
This work was supported by the Applications Driving Architectures (ADA) Research Center, a JUMP Center co-sponsored by SRC and DARPA, and IBM-ILLINOIS Center for Cognitive Computing Systems Research (C3SR) -- a research collaboration as part of the IBM AI Horizon Network.

\bibliographystyle{IEEEtran}
\bibliography{IEEEabrv,IEEEexample}

\end{document}